\documentclass[lettersize,journal]{IEEEtran}
\usepackage{amsmath,amsfonts}
\usepackage{algorithmic}
\usepackage{array}
\usepackage{subfigure}
\usepackage[caption=false,font=normalsize,labelfont=sf,textfont=sf]{subfig}
\usepackage{textcomp}
\usepackage{bm}
\usepackage{cite}
\usepackage{amsthm}
\usepackage{amsmath}
\usepackage{amssymb}
\usepackage{color}

\usepackage{stfloats}
\usepackage{url}
\usepackage{verbatim}

\usepackage{graphicx}

\def\BibTeX{{\rm B\kern-.05em{\sc i\kern-.025em b}\kern-.08em
    T\kern-.1667em\lower.7ex\hbox{E}\kern-.125emX}}
\usepackage{balance}
\usepackage{soul}

\usepackage{colortbl}
\usepackage{booktabs} 
\usepackage[table,xcdraw]{xcolor}

\begin{document}
\title{
LNN-powered Fluid Antenna Multiple Access}
\author{
Pedro D. Alvim, Hugerles S. Silva, Ugo S. Dias, Osamah S. Badarneh, Felipe A. P. Figueiredo and Rausley A. A. de Souza

 \thanks{P. D. Alvim, H. S. Silva, and U. S. Dias are with the Electrical Engineering Department, University of Brasília (UnB), Federal District, Brazil. (pedro.alvim@redes.unb.br, hugerles.silva@unb.br, udias@unb.br). O. S. Badarneh is with the Department of Electrical Engineering,
German Jordanian University, Amman 11180, Jordan, and also with the
Communications Engineering Department, Princess Sumaya University for
Technology, Amman 11941, Jordan (e-mail: osamah.badarneh@gju.edu.jo). 
F. A. P. Figueiredo and R. A. A. de Souza are with the National Institute of Telecommunications (Inatel), Santa Rita do Sapucaí 37536-001, Brazil. (felipe.figueiredo@inatel.br, rausley@inatel.br). 
 This work was partially supported by CNPq (311470/2021-1 and 403827/2021-3),  by the projects XGM-AFCCT-2024-2-5-1, XGM-FCRH-2024-2-1-1, and XGM-AFCCT-2024-9-1-1 supported by xGMobile - EMBRAPII-Inatel Competence Center on 5G and 6G Networks, with financial resources from the PPI IoT/Manufatura 4.0 from MCTI, 052/2023, signed with EMBRAPII,  by RNP, with resources from MCTIC, 01245.020548/2021-07, under the Brazil 6G project of the Radiocommunication Reference Center (Centro de Referência em Radiocomunicações - CRR) of the National Institute of Telecommunications (Instituto Nacional de Telecomunicações - Inatel), and by Fapemig (PPE-00124-23, APQ-04523-23, APQ-05305-23 and APQ-03162-24).}

}

\markboth{Submitted to IEEE COMMUNICATIONS LETTERS, APRIL~2025}%
{How to Use the IEEEtran \LaTeX \ Templates}

\maketitle

\begin{abstract}
Fluid antenna systems represent an innovative approach in wireless communication, recently applied in multiple access to optimize the signal-to-interference-plus-noise ratio through port selection. 
This letter frames the port selection problem as a multi-label classification task for the first time, improving best-port selection with limited port observations. 
We address this challenge by leveraging liquid neural networks (LNNs) to predict the optimal port under emerging fluid antenna multiple access scenarios alongside a more general $\alpha$-$\mu$ fading model.
We also apply hyperparameter optimization to refine LNN architectures for different observation scenarios. 
Our approach yields lower outage probability values than existing methods.
\end{abstract}
   

\section{Introduction}

\IEEEPARstart{I}{n} recent years, various advanced technologies, including multiple input multiple output (MIMO) systems, non-orthogonal multiple access (NOMA), reconfigurable intelligent surfaces (RIS), and more recently fluid antenna system (FAS), have been introduced and developed to improve the radio interface for fifth generation~(5G) and beyond (5G+/6G).

FASs have attracted considerable interest~\cite{Wongbruce,Wong2020,Wong,Wong2023c} as they offer a solution to the spatial limitations of MIMO by dynamically adjusting antenna positions within a limited area to improve performance \cite{Wongbruce,Wong2020,Wong}. 
FASs are discussed in \cite{Wongbruce}, and their performance limits are analyzed in~\cite{Wong2020}, which presents expressions for important statistics and metrics. 
Recently, FAS has been adapted for multi-user access, known as fluid antenna multiple access~(FAMA)~\cite{Wong2022,Wongslowfama,Shah}, which exploits different user-fading envelopes to enable efficient multi-user connections with low processing demands \cite{Wong2022,Wongslowfama,Shah}.
FAMA provides very high data speed and low latency, as well as supports high link density, high energy and spectral efficiency, high reliability, and high mobility~\cite{Shah}. 
These features are important to achieve the 5G+/6G requirements and enable FAMA to be combined with technologies such as deep learning, artificial intelligence, MIMO, NOMA, RIS, and terahertz communication. 
This combination allows FAMA to be applied to real-world solutions such as the smart grid, the Internet of Things, industrial internet, smart cities, smart transportation, and smart health. 
Several challenges and solutions for real scenarios in which FAMA can be applied are described in~\cite{Shah}.

In FAS and FAMA, it is assumed that the system can select the port with the best signal-to-noise ratio (SNR) and signal-to-interference-plus-noise ratio (SINR), respectively \cite{Wong2022,Wongslowfama}. 
However, with a large number of ports \(N\), obtaining channel state information (CSI) for all ports in real-time within the coherence time is impractical.
Therefore, a critical challenge in FAS is the port selection problem, in which the system must determine the optimal antenna port to maximize signal quality. 
This problem is even more challenging in FAMA, which must maximize signal quality and minimize interference across multiple simultaneously served users with distinct fading profiles. 
Due to the difficulty of obtaining signal quality or CSI for all ports, advanced techniques, such as machine learning, are essential to efficiently select the best port based on limited observations (i.e., observing only a small percentage of ports). 
Solving this problem is key to maximizing FAS and FAMA performance in next-generation networks.

In the literature, the port selection problem in FAS and FAMA has been investigated in several studies \cite{Chai,ChecarAutores,Waqar,Zhang,Eskandari}. 
In~\cite{Chai}, machine learning (ML) methods such as long short-term memory (LSTM) and a novel algorithm called smart, predict, and optimize (SPO) are used to predict the optimal port in FAS by observing only a fraction of ports. 
The results in~\cite{Chai} show a reduction in outage probability (OP), even with only $10\%$ of the ports being observed. 
In~\cite{Chai}, the correlation model adopted by the authors has the disadvantage of requiring a reference port, which is a simplistic and unrealistic assumption.
In~\cite{ChecarAutores}, the port selection problem in FAS is revisited, also considering the use of LSTM and SPO algorithms, but adopting a more realistic correlation model \cite{KAIKITNOVO}, where any port can be a reference for another one. 
LSTM is applied under a slow FAMA (s-FAMA) scenario in~\cite{Waqar}, considering a more accurate model to emulate spatial correlation across the ports. 
The aforementioned work obtained good results with only $25\%$ of the ports observed.
In~\cite{Zhang}, an LSTM-based learning approach is considered to estimate and predict the CSI of the port for fast selection in FAS, exploring the temporal and spatial correlation of the channels. 
In~\cite{Eskandari}, the SINR is inferred from only a few observations in fluid antennas using a conditional generative adversarial network (cGAN), where OP improvements are obtained. 
The port selection problem is framed as a regression task in all these related works.
Moreover, they all adopted correlated Rayleigh fading.

In this letter, for the first time, we frame the port selection problem as a multi-label classification (MLC) task under emerging and distinct FAMA scenarios, leveraging liquid neural networks (LNNs) alongside a more general channel fading model, the $\alpha$-$\mu$. 
MLC allows the ML model to focus solely on predicting the best port rather than estimating the port's CSI or SINR/SNR value, simplifying the problem \cite{optuna}. 
Regarding the scenarios, we first analyze a downlink communication system in which the base station (BS) transmits messages to users, and only one port is activated at a time \cite{Wong2022,Wongslowfama}. 
We also investigated the recently proposed scenario in \cite{Lai}, but in the context of FAMA, where multiple ports can be activated for a subsequent signal combination to improve receiver performance. 
In this case, given $N$ ports in FAMA, only the best $M$ ports are selected before performing the maximum ratio combining~(MRC) of the branches that contain the signals received from the selected ports.

The recent fully port correlation model presented in~\cite{Khammassi} is adopted. 
Furthermore, we employ a hyperparameter optimization framework to find optimal LNN-based architectures for each number of observed ports. 
Our results are compared with those in the literature, demonstrating better capabilities in predicting the best port and thus achieving lowered OP values. 
We consider LNNs due to their superior capabilities in real-time modeling and adaptation to dynamic channel changes and their ability to efficiently capture temporal and spatial dependencies that allow predicting the port that maximizes SINR~\cite{Zhu}. 
This approach offers an effective and scalable alternative to traditional techniques. 
It mitigates the challenges of obtaining signal quality metrics or CSI for all ports and improves the feasibility and performance of FAMA systems.
We assume the $\alpha$-$\mu$ fading model as the channel between the BS and the users since the aforementioned fading model is generalist, encompassing several fading models as particular cases; it is supported by experimental results and characterized by physical parameters.
The suitability of the $\alpha$-$\mu$ fading channel model for FAS is evidenced in~\cite{Alvim}.

In a nutshell, the main and original contributions of this work can be described brieﬂy as follows: i) This letter frames the port selection problem as a MLC task, improving best-port selection with limited port observations (i.e., the models predict the indices of the ports more likely to have the highest SNIR values, eliminating the need to estimate continuous SNIR values); ii) We leverage LNNs to predict the optimal port under FAMA scenarios alongside the $\alpha$-$\mu$ fading model, iii) We also apply hyperparameter optimization to refine LNN architectures for different observation scenarios, and iv) Our approach yields lower OP values than existing methods.


\section{System Model}\label{system}

\subsection{FAMA}

A downlink communication system model is considered, where a BS transmits messages to $U$ users \cite{Wong2022}. 
The BS is equipped with $U$ antennas, each dedicated to transmitting signals to a specific user in the downlink. 
On the user side, each device is equipped with an $N$-port FAS. The system adopts a feasible and practical implementation of FAMA, commonly referred to in the literature as s-FAMA~\cite{Wongslowfama}.

\begin{figure}
    \begin{center}
    \includegraphics[scale=0.85]{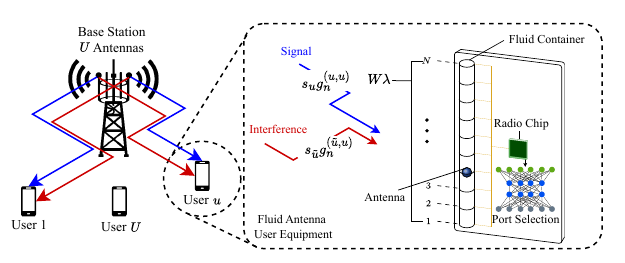}   
    \caption{A possible architecture for FAMA.}
\label{fig:fig2}
\end{center}
\end{figure}

In this approach, the selected antenna port remains fixed until the channel conditions change. 
Notably, as highlighted in \cite{Wong2022}, FAMA simplifies multiple access by reducing it to a port selection task on the user side, eliminating the need for pre-coding or CSI at the BS. A conceptual architecture for FAMA is illustrated in Fig.~\ref{fig:fig2}, which depicts a network consisting of a BS and user equipment equipped with a fluid antenna.

In this system, the fluid antenna is an electronically reconfigurable structure based on pixel technology. It comprises a linear array of predefined positions, referred to as ports, where the radiating element is dynamically switched to optimize a reception metric. In Fig.~\ref{fig:fig2}, $W$ represents the normalized size of the fluid antenna, while $\lambda$ denotes the propagation wavelength.

The signal received at the $n$-th port of the $u$-th user equipped with a fluid antenna is given by~\cite[Eq. (1)]{Wongslowfama}
$y^{(u)}_{n} = s_{u} g^{(u,u)}_{n}+ \sum _{\substack{\tilde {u}=1\\ \tilde{u}\ne u}}^{U} s_{\tilde {u}}   g^{(\tilde{u},u)}_{n}          + \eta_{n}^{(u)}$, 
where $s_u$ denotes the symbol transmitted to the $u$-th user with power $\sigma_\text{s}^{2}$, $g^{(u,u)}_{n}$ denotes the channel gain between the $u$-th BS antenna and the $n$-th port of the $u$-th user, $g^{(\tilde{u},u)}_{n}$ denotes the channel gain between the $\tilde{u}$-th BS antenna and the $n$-th port of the $u$-th user, acting as interference on this port, and $\eta^{(u)}_n$ represents the additive white Gaussian noise (AWGN) with power $\sigma_{\eta}^{2}$. 
The channel gains are assumed to follow the $\alpha$-$\mu$ distribution and the correlation model presented in~\cite{Khammassi} is adopted.

In FAMA, the objective is to maximize the instantaneous SINR across all $N$ ports of the $u$-th user, which is given by~\cite[Eq. (7)]{Waqar}
    \begin{equation}\label{eq:SINR}
        \gamma_{n}^{(u)} = \frac{\sigma_\text{s}^{2}|g_{n}^{(u,u)}|^{2}}{\sigma_\text{s}^{2}\sum _{\substack{\tilde {u}=1\\ \tilde {u}\ne u}}^{U} |g^{(\tilde{u},u)}_{n}|^2 + \sigma_{\eta}^{2}}.
    \end{equation}

In the scenario previously described, it should be noted that only one port is activated at a time. 
However, we also analyze the recent approach proposed in~\cite{Lai}, where multiple ports can be activated for a subsequent signal combination to improve the receiver performance. 
In this scenario, only the best $M$ out of $N$ available ports are selected and combined using MRC.
In~\cite{Lai}, the port selection problem has not been investigated in FAS or FAMA contexts. 
Therefore, this is another original contribution of this work.

\subsection{Port Selection} \label{system2}

In FAS or FAMA systems, the port with the highest SNR or SINR is always assumed to be selected. 
This search process is straightforward if the receiver knows all the channel gains. 
However, the number of ports in FAS or FAMA is considerable. 
Therefore, estimating the channel gain for all ports can be impractical from a computational and cost perspective.

In the scenario where only a small subset of ports, $\cal{M}$, is observed, the port selection problem becomes~\cite{Chai}
    \begin{equation}\label{eq:selecaoPortas2}
        \mathcal{K} = \underset{ n \in \mathcal{N} }{\arg \; K \; \max}  \{\{\gamma_{n}^{(u)}\}_{n \in \cal{M}}, \{\tilde{\gamma}_{n}^{(u)}\}_{n \in \cal{Z}}\},
    \end{equation}
in which $\mathcal{K}$ is the set of indices of the optimal ports, $K$ indicates the number of indices of ports with the highest values (i.e., highest SINR) to select out of the set of all available ports, $\mathcal{N}$, $\cal{M}$ represents the set of indices of port values (i.e., SINR) known by direct observation, $\cal{Z}$ represents the set of indices of port values obtained by other means (e.g., ML models), and $\tilde{\gamma}_{n}^{(u)}$ represents the port value acquired for the $n$-th non-observed port. 
Note that $ \mathcal{N} = \mathcal{M} \; \cup \; \mathcal{Z}$ and that $K > 1$ is used for implementing MRC.

We introduce a novel approach to the port selection problem by treating it as a classification task, bypassing the need to predict the actual SINR value, i.e., a regression task. 
Instead, we frame the port selection problem as an MLC task, where the model predicts the indices of the $M$ ports (i.e., classes) more likely to have the highest SINR values. 
Once the model predicts indices, not the actual SINR values, the FAS must observe the ports corresponding to the $K$ optimal indices. 
\subsection{Performance Evaluation Metric}
The performance evaluation of FAMA is conducted using the OP metric, denoted as $P_{\rm out}$. 
Note that $P_{\rm out}$ represents the probability that the SINR falls below a threshold $\gamma_{\rm th}$. 
In FAMA, $P_{\rm out}$ for a user $u$ employing the MRC technique can be calculated using~(\ref{eq:SINR}) as 
$P_{\rm out} = \mathrm{Prob} \left( \gamma_{\text{MRC}}^{(u)} < \gamma_{\rm th} \right)$,
where
{\small
\begin{equation}
    \gamma_{\text{MRC}}^{(u)} = \frac{\sigma_\text{s}^{2} \left( \sum_{n \in \mathcal{K}} |g_{n}^{(u,u)}|^{2} \right)^2}{\sigma_\text{s}^{2} \sum_{n \in \mathcal{K}} | \sum _{\substack{\tilde {u}=1\\ \tilde {u}\ne u}}^{U} g_{n}^{*(u,u)}g^{(\tilde{u},u)}_{n}|^2 + \sigma_{\eta}^{2}\sum_{n \in \mathcal{K}} |g_{n}^{(u,u)}|^{2}}.
\end{equation}
}
\vspace{-0.5cm}
\section{Deep Learning-Based Port Selection}\label{IAselecao}
\subsection{LSTM}
LSTM is a recurrent neural network architecture widely applied across various practical contexts. 
Due to its memory retention characteristics, it is commonly used to model time series data. 
However, it has been employed to model spatially correlated signals at the antenna ports in FAS and FAMA.
In \cite{Chai,Waqar}, LSTM performs a regression task predicting channel conditions based on a subset of observed ports. 
It enables port selection by estimating the gains of unobserved ports through its predictive capability.
Using supervised learning and analyzing the port selection problem as a regression task, the LSTM model is trained to predict the values of the $N$ ports, considering the mean squared error (MSE) as the loss function.
\subsection{LNN}
LNNs are first introduced in \cite{hasani2021} to enhance the stability of dynamic time series through a specialized layer known as the liquid layer, which incorporates solvers for ordinary differential equations (ODEs). 
Unlike traditional recurrent neural networks, such as LSTM, LNNs feature an adaptive structure that modifies dynamically in response to new data, enabling more efficient processing and improved energy utilization. 
This adaptability allows LNNs to be trained faster and with fewer computational resources than conventional recurrent neural networks, such as LSTM networks.

At the core of LNNs is the liquid time constant (LTC) layer, which relies on ODEs to adjust dynamically to input data changes over time~\cite{hasani2021}. 
Although these equations increase computational complexity, they allow the network to adapt to evolving data patterns. 
LTCs have been applied in various fields, including time series forecasting for training autonomous vehicles, channel estimation, and vehicle traffic modeling. 
In this work, we employ LNNs to predict spatially correlated channels.
\subsection{Optuna-aided Optimization}
This work employs a hyperparameter optimization framework, Optuna, to identify optimal liquid-layer-based neural network architectures tailored to different numbers of observed ports \cite{optuna}.
This framework conducts multiple training sessions across various neural network architectures.
The following hyperparameters are adjusted to minimize the defined classification error for different numbers of observed ports: application of preprocessing techniques, such as principal component analysis (PCA) and scaling; number and types of layers, including liquid (LNN), convolutional (CNN) and dense (DNN) layers; number of filters, cells, and units in the convolutional, liquid, and dense layers, respectively; choice of optimizers; loss functions; and learning rate. 
Consequently, optimized architectures are determined for each configuration of observed ports.

As mentioned, the port selection problem is framed as an MLC task, where the LNN-based models are trained to predict the set of $M$ port indices with the highest probability values. 
$M$ is a hyperparameter that must be optimized as well. 
The models' performance is assessed using F1-score or binary cross-entropy as the loss function.
\section{Results and Remarks}\label{resultados}
\subsection{Parameter Configuration}
The data considered in this study are generated using MATLAB software, assuming the $\alpha$-$\mu$ channel's envelope gains.
The neural network models based on LSTM and LNN layers are implemented in Python using the TensorFlow framework. 
The generated data are divided into two parts: the input attributes, which are the values of the subset of observed ports, and labels, which are the output values of the models.
In the case of LSTM-based models, the labels correspond to the raw data generated, i.e., the actual SINR values read across all $N$ ports, since the port selection problem is addressed as a regression task.
For the LNN-based models, where MLC is assumed, the labels corresponding to the $M$ highest values among the $N$ ports are represented by values equal to $1$, meaning their probability of presenting one of the $M$ highest SINR values is one. 
In contrast, the other labels (i.e., $N-M$) are represented by values equal to $0$, meaning their probability of presenting one of the $M$ highest SINR values is zero. 
The dataset for both cases is divided into training, validation, and testing sets, with percentages of $70\%$, $15\%$, and $15\%$, respectively. 
We assume that the samples in the validation and testing sets are unseen by the trained model, meaning the model is not presented with them during training.

Our study considers a linear structure\footnote{As future work, we aim to analyze the port selection problem considering a two-dimensional FAS.} of the fluid antenna and that the average SNR at each UE is set to $40$ dB.
Furthermore, the observed ports are chosen to be uniformly distributed over $ W\lambda$ to explore the correlation effect more effectively.
It should be mentioned that the results obtained with the LSTM-based architecture presented in~\cite{Waqar} are used for comparison with our results. 
Additionally, the optimal hyperparameters for the LNN-based architectures for each observed port count are available in a public repository, along with the source code needed to reproduce our findings \footnote{The code is available at: https://github.com/AlvimPedro/LNN-FAMA}. 
In the mentioned repository, we also provide a discussion of the computational complexity of the model and its convergence, as well as the computational cost under different settings.

\begin{table}
\centering
\caption{OP as a function of number of classes and observed ports.}
\resizebox{\columnwidth}{!}{ 
\begin{tabular}{|c|c|c|c|c|c|c|c|}
\hline
\rowcolor[HTML]{EFEFEF} 
& \multicolumn{6}{|c|}{\textbf{Number of classes, $M$}} \\ \hline
\begin{tabular}[c]{@{}c@{}}\textbf{Number of}\\ \textbf{observed}\\\textbf{ports}\end{tabular} & \textbf{1} & \textbf{2}     & \textbf{3}  & \textbf{4}    & \textbf{5}      & \textbf{10} \\ \hline
$5$  & $0.1904$   & $0.1884$ & \cellcolor[HTML]{C1FFC1}$0.1881$ & $0.1882$ & $0.1890$ & $0.1901$ \\ \hline
$6$  & $0.1543$   & $0.1499$ & \cellcolor[HTML]{C1FFC1}$0.1494$ & $0.1500$ & $0.1507$ & $0.1502$ \\ \hline
$7$  & $0.1107$   & $0.1097$ & \cellcolor[HTML]{C1FFC1}$0.1089$ & $0.1097$ & $0.1091$ & $0.1089$ \\ \hline
$8$  & $0.0839$   & $0.0834$ & \cellcolor[HTML]{C1FFC1}$0.0834$ & $0.0834$ & $0.0842$ & $0.0863$ \\ \hline
$9$  & $0.0703$   & $0.0690$ & \cellcolor[HTML]{C1FFC1}$0.0686$ &  $0.0686$ & $0.0690$ & $0.0709$ \\ \hline
$10$ & $0.0561$   & $0.0548$ & \cellcolor[HTML]{C1FFC1}$0.0544$ & $0.0548$ & $0.0554$ & $0.0549$ \\ \hline
$11$ & $0.0457$   & $0.0452$ & \cellcolor[HTML]{C1FFC1}$0.0446$ & $0.0450$ & $0.0474$ & $0.0448$ \\ \hline
$13$ & $0.0287$   & $0.0282$ & \cellcolor[HTML]{C1FFC1}$0.0281$ & $0.0281$ & $0.0281$ & $0.0293$ \\ \hline
$15$ & $0.0212$   & $0.0211$ & \cellcolor[HTML]{C1FFC1}$0.0211$ & $0.0212$ & $0.0218$ & $0.0223$ \\ \hline
\end{tabular}
}
\label{tab:m_classes}
\end{table}
\begin{figure}
\centering
    \includegraphics[trim={0 0 0  0cm},scale=0.5]{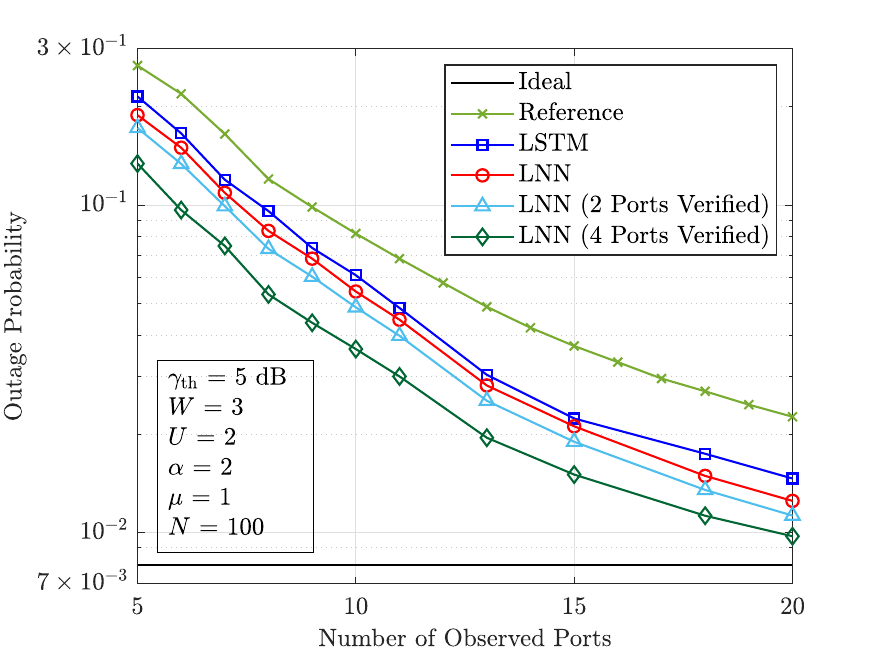}   
    \caption{{OP curves as a function of the number of observed ports for s-FAMA.}}
\label{fig:OPFAMA}
\end{figure}

\begin{figure}
    \begin{center}
    \includegraphics[trim={0 0 0 1cm},scale=0.5]{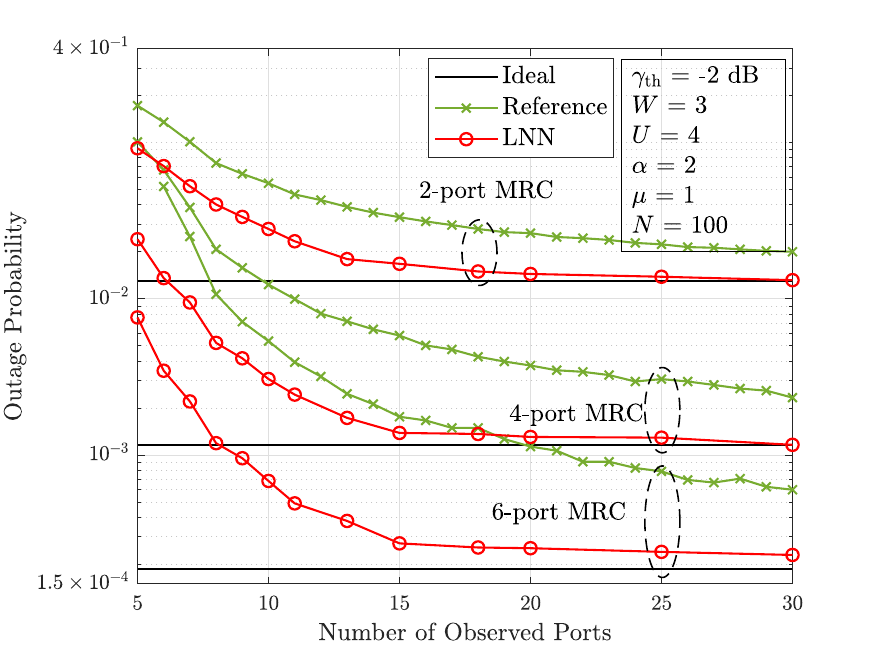}   
    \caption{OP curves as a function of the number of observed ports, under s-FAMA, with different values for the MRC  branches.}
\label{fig:MRCop}
\end{center}
\end{figure}

\subsection{Selecting the Number of Classes, $M$}
Table~\ref{tab:m_classes} compares the OP attained when Optuna is employed to find optimized LNN-based architectures for each number of classes, $M$, and observed ports with the parameters listed in the bottom-left box of Fig.~\ref{fig:OPFAMA}. 
Then, using each optimized architecture, the port selected for calculating the OP is the one meeting \eqref{eq:selecaoPortas2} when $K=1$. 
The table shows that $M = 3$ is the best number of classes since it produces the lowest OP across all considered numbers of observed ports. 
Therefore, we make $M=3$ henceforward for all subsequent results.
The more labels, $M$, an instance has, the higher the chance the model will make an error in at least one of them, increasing the total loss. Moreover, predicting multiple labels requires the model to learn intricate dependencies between labels. As $M$ grows, the decision boundary becomes more complex in a high-dimensional space.


\subsection{Performance Analysis}
As a function of the number of observed ports, OP curves are presented in Fig.~\ref{fig:OPFAMA} for s-FAMA, considering the proposed optimized LNN-based architectures and the LSTM-based one presented in~\cite{Waqar}.  
The ideal-case curve corresponds to the situation where the gain of all ports is available for the subsequent selection of the best one. 
Note that this curve can be interpreted as an upper bound.  
The reference-case curve refers to the case where the one-best port is selected only among the set of observed ones, meaning that the highest value among the observed ports is chosen.
The OP decreases as the number of observed ports increases.  

The optimized LNN-based models perform better than the LSTM-based model presented in the literature~\cite{Waqar}. 
For $20\%$ of the observed ports, the LNN performance is already close to the ideal case, tending to it as the number of ports increases.
The dark-blue and red curves used the methodology adopted in~\cite{Waqar}, which selects the port with the highest SINR among those observed and the one indicated by the model (i.e., the port with the index predicted by the model as having the highest SINR), as expressed by \eqref{eq:selecaoPortas2}. 

Additionally, we extended that methodology to include a subset of ports indicated by the model as having the highest probabilities in the port selection problem. 
This modified methodology selects the port with the highest SINR among those observed and the subset of ports indicated by the model. 
The results of this selection methodology are illustrated in Fig.~\ref{fig:OPFAMA} by the light-blue and dark-green curves, where the SINR values of a subset of $2$ and $4$ ports indicated by the model are also considered for the port selection, respectively. 
As can be seen, this approach improves the system performance but at the cost of higher computational complexity.

Fig.~\ref{fig:MRCop} presents OP curves as a function of the number of observed ports for different numbers of combined ports. 
We consider the scenario in which multiple ports can be activated for subsequent signal combinations to enhance the receiver's performance. 
In this case, $K$ is set to a value greater than $1$, i.e., $2$, $4$, and $6$ ports, respectively.
In this case, s-FAMA and the MRC technique are combined. 
We chose $\gamma_{\text{th}} = -2$~dB to analyze the system behavior under practical OP values.
As expected, for a given number of observed ports, the OP decreases as the number of combined ports increases. 
As the number of combined ports increases, the gap between the reference curve and the curve predicted by the LNN model widens.
This behavior is a direct consequence of combining more ports since the probability of getting a port with SINR above the threshold is higher. 
Notably, from $30$ observed ports, the proposed LNN-based model’s predictions align with the ideal-case curve across the scenarios considering $2$ and $4$ combined ports. 
However, the increased system robustness concerning OP has the disadvantage of increased complexity and cost, as more radio frequency chains and higher processing of received signals are required.
\begin{figure}
    \begin{center}
    \subfigure[]
    {
        \includegraphics[trim={0 0 0 1cm},scale=0.5]{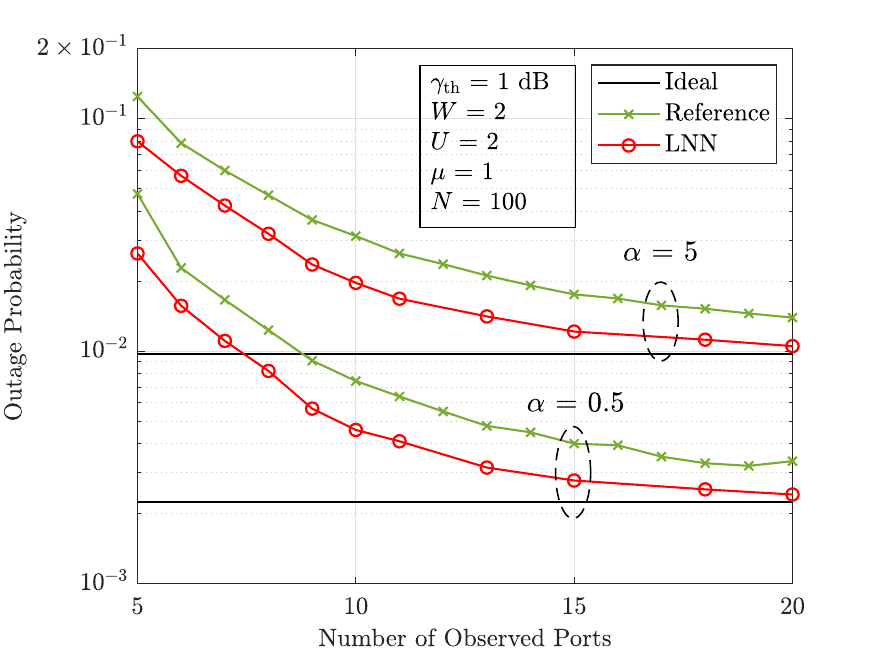} 
    }\\
    \subfigure[]
    {
        \includegraphics[trim={0 0 0 0.6cm},scale=0.5]{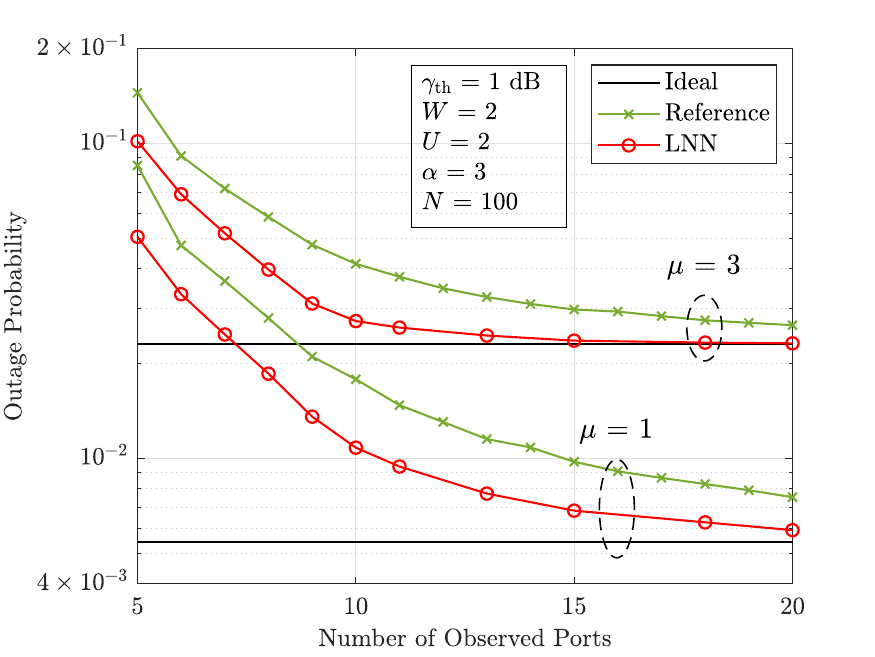} 
    }
\caption{OP curves as a function of the number of observed ports, under s-FAMA, for different values of (a) $\alpha$ and (b) $\mu$.}
\label{fig:alphas}
    \end{center}
\end{figure}

Figs. \ref{fig:alphas}(a) and (b) highlight the impact of channel non-linearity and fading intensity on the OP, respectively.
For $\mu=1$, the Weibull fading model can be obtained as a particular case of the study proposed in this work. 
To the best of the authors' knowledge, the literature does not present results for the port selection under FAMA considering the Weibull model. 
The following insights are perceived where the OP improves: (i) as the number of observed ports increases; (ii) as $\alpha$ and/or $\mu$ decreases.
In fact, to provide multiple access, the FAMA system utilizes the deep fades of interference, which occur more frequently in harsh environments (i.e., the FAMA system has more opportunities to exploit signal envelope variations and peaks).
Note that the LNN can predict the optimal port even for different values of $\alpha$ and $\mu$, yielding near-ideal results comparable to those in other scenarios for $20$ observed ports.
This adaptability shows the capability of the LNN to perform effectively across various conditions.
%
%


\begin{thebibliography}{}
%
 \bibitem{Wongbruce} K. -K. Wong et al., ``Fluid Antenna System for 6G: When Bruce Lee Inspires Wireless Communications,'' \textit{Electron. Lett.}, vol. 56, no. 24, pp. 1288-1290, Nov. 2020.

 \bibitem{Wong2020} K. K. Wong et al., ``Performance Limits of Fluid Antenna Systems,'' \textit{IEEE Commun. Lett.}, vol. 24, no. 11, pp. 2469-2472, Nov. 2020.

\bibitem{Wong} K. -K. Wong et al., ``Fluid Antenna Systems,'' \textit{IEEE Trans. Wirel. Commun.}, vol. 20, no. 3, pp. 1950-1962, Mar. 2021.



\bibitem{Wong2023c} K. -K. Wong, K. -F. Tong and C. -B. Chae, ``Fluid Antenna System—Part II: Research Opportunities,'' \textit{IEEE Commun. Lett.}, vol. 27, no. 8, pp. 1924-1928, Aug. 2023




 \bibitem{Wong2022} K. -K. Wong and K. -F. Tong, ``Fluid Antenna Multiple Access,'' \textit{IEEE Trans. Wirel. Commun.}, vol. 21, no. 7, pp. 4801-4815, Jul. 2022.




\bibitem{Wongslowfama} K. -K. Wong et al., ``Slow Fluid Antenna Multiple Access,'' \textit{IEEE Trans. Commun.}, vol. 71, no. 5, pp. 2831-2846, May 2023.


\bibitem{Shah} A. F. M. S. Shah etal., ``A Survey on Fluid Antenna Multiple Access for 6G: A New Multiple Access Technology That Provides Great Diversity in a Small Space,''\textit{ IEEE Access}, vol. 12, pp. 88410-88425, Jun. 2024.

\bibitem{Chai} Z. Chai et al., ``Port Selection for Fluid Antenna Systems,'' \textit{IEEE Commun. Lett.}, vol. 26, no. 5, pp. 1180-1184, May 2022.


\bibitem{ChecarAutores} Z. Chai et al., ``Performance of Machine Learning Aided Fluid Antenna System with Improved Spatial Correlation Model,'' in Proc. of the \textit{IEEE 6GNet}, pp. 1-6, 2022.


 \bibitem{Waqar} N. Waqar et al., ``Deep Learning Enabled Slow Fluid Antenna Multiple Access,'' \textit{IEEE Commun. Lett.}, vol. 27, no. 3, pp. 861-865, Mar. 2023.

\bibitem{Zhang} S. Zhang et al., ``Fast Port Selection using Temporal and Spatial Correlation for Fluid Antenna Systems,'' in Proc. of \textit{IEEE Statistical Signal Processing Workshop}, pp. 95-99, 2023.

\bibitem{Eskandari} M. Eskandari et al., ``cGAN-Based Slow Fluid Antenna Multiple Access,'' \textit{IEEE Wirel. Commun. Lett.}, vol. 13, no. 10, pp. 2907-2911, Oct. 2024.


 \bibitem{KAIKITNOVO}  K. -K. Wong et al., ``Closed-Form Expressions for Spatial Correlation Parameters for Performance Analysis of Fluid Antenna Systems,'' \textit{Electron. Lett.}, vol. 58, no. 11, pp. 453-457, May 2022.



 \bibitem{Lai} X. Lai et al., ``On Performance of Fluid Antenna System using Maximum Ratio Combining'', \textit{IEEE Commun. Lett.}, vol. 28, no. 2, pp. 402-406, Fev. 2024.


\bibitem{Khammassi} M. Khammassi, A. Kammoun and M. -S. Alouini, ``A New Analytical Approximation of the Fluid Antenna System Channel,'' \textit{IEEE Trans. Wirel. Commun.}, vol. 22, no. 12, pp. 8843-8858, Dec. 2023.

\bibitem{Zhu} F. Zhu et al, ``Robust Continuous-Time Beam Tracking with Liquid Neural Network, \textit{ArXiv}, pp. 1-6, Aug. 2024.

\bibitem{Alvim} P. D. Alvim et al., ``On the Performance of Fluid Antenna Systems Under $\alpha$-$\mu$ Fading Channels,'' \textit{IEEE Wirel. Commun. Lett.}, vol. 13, no. 1, pp. 108-112, Jan. 2024.

\bibitem{hasani2021} R. Hasani et al., ``Liquid Time-Constant Networks'', in Proc. of \textit{AAAI Conference on Artificial Intelligence}, vol. 35, no. 9, 2021.

\bibitem{optuna} T. Akiba et al., ``Optuna: A Next-Generation Hyperparameter Optimization Framework'', in Proc. of  \textit{25th ACM SIGKDD International Conference on Knowledge Discovery \& Data Mining}, 2019.


  
\end{thebibliography}
\end{document}